\documentclass{PoS}

\def\be{\begin{equation}}
\def\ee{\end{equation}}
\def\bea{\begin{eqnarray}}
\def\eea{\end{eqnarray}}

\title{Information Retention by Stringy Black Holes}

\ShortTitle{Stringy Black Holes Retain Information}

\author{John Ellis, Nick E. Mavromatos\thanks{Work supported in part by the European Research Council via the Advanced Investigator
Grant 267352 and by the UK STFC via the Research Grant ST/L000326/1.}\\
        Theoretical Particle Physics and Cosmology Group, Department of Physics, \\
King's College London, Strand, London WC2R 2LS, U.K; \\
Theory Division, Physics Department, CERN, CH 1211 Geneva 23, Switzerland\\
        E-mail: \email{John.Ellis@cern.ch, nikolaos.mavromatos@kcl.ac.uk}}

\author{D. V. Nanopoulos\thanks{Work supported in part by the DOE Research Grant DE-FG02-13ER42020.}\\
        George P. and Cynthia W. Mitchell Institute for Fundamental Physics and Astronomy,\\
Texas A\&M University, College Station, TX 77843, USA; \\
Astroparticle Physics Group, Houston Advanced Research Center (HARC), \\
Mitchell Campus, Woodlands, TX 77381, USA;\\
Academy of Athens, Division of Natural Sciences, Athens 10679, Greece\\
        E-mail: \email{dimitri@physics.tamu.edu}
           ~\\
           ~\\
           {\tt  KCL-PH-TH/2015-51, LCTS/2015-38, CERN-PH-TH/2015-258, \\ ACT-09-15, MI-TH-1539}  }

\abstract{Building upon our previous work on two-dimensional stringy black holes and its
extension to spherically-symmetric four-dimensional stringy black holes, we
show how the latter retain information. A key r\^ole is played by an
infinite-dimensional $W_\infty$ symmetry that preserves the area of
an isolated black-hole horizon and hence its entropy. The exactly-marginal conformal world-sheet operator 
representing a massless stringy particle interacting with the 
black hole necessarily includes a contribution from $W_\infty$ generators
in its vertex function. This admixture manifests the transfer of
information between the string black hole and external particles.
We discuss different manifestations of $W_\infty$ symmetry
in black-hole physics and the connections between them.}

\FullConference{18th International Conference From the Planck Scale to the Electroweak Scale \\
		 25-29 May 2015\\
		 Ioannina, Greece}

\begin{document}

\section{Introduction}

The black-hole information problem requires no introduction. Decades ago,
Bekenstein~\cite{bek} and Hawking~\cite{hawk} discovered that four-dimensional black holes have
thermodynamical properties such as temperature and non-zero entropy
corresponding to a mixed quantum-mechanical
state. Hawking, in particular, then argued~\cite{hawk2} that information would be lost across
the black-hole horizon, giving rise to a transition from a pure to a mixed state.

The advent of string theory, and in particular Witten's construction~\cite{witt} of a
two-dimensional black hole solution using an SU(1,1)/U(1) coset structure~\cite{horava,Chlyk},
coupled with dualities~\cite{KN} and
followed by the construction of four-dimensional stringy black holes using
D-branes~\cite{dbranebh,zanon}, provided opportunities to probe the black-hole information problem
in an explicit theoretical laboratory. We have argued~\cite{emn1} that two-dimensional
black holes carry an infinite set of `hairy' $W$ quantum numbers that preserve {\it in
principle} the lost information, though {\it in practice} this information is
inaccessible. We have also argued that these observations can be extended
to spherically-symmetric four-dimensional black holes~\cite{emn2}, whose horizon
geometry is encoded in a similar SU(1,1)/U(1) coset structure, accompanied
by a similar infinite-dimensional $W$ symmetry and an associated infinite set
of `hair' that is measurable in principle~\cite{wmeasure}.
 
Constructions using D-branes provided explicit examples of four-dimensional
black-hole solutions whose microstates could be counted~\cite{dbranebh}, giving numbers
consistent with the Bekenstein-Hawking entropy and
suggesting that indeed the `lost' information could {\it in principle} be retained.
However, there still remained the issues how the information was transferred
to and from these microstates, in what form it was stored,
and whether the information transferred into the microstates could 
{\it in practice} be extracted, or whether it was in reality lost.

An interesting approach to addressing these issues has recently been taken
in a series of papers by Strominger and collaborators~\cite{strominger,strominger2}. They have shown that
spherically-symmetric four-dimensional black holes carry an infinite set
of gravitational `hair' associated with BBMS supertranslations~\cite{bms,bms2} on the retarded
null infinity $\cal{I}^+$, corresponding to vacua that differ by the addition of
soft gravitons and could be measured via the gravitational memory effect.
They also found an infinite set of inequivalent electromagnetic gauge
configurations corresponding to electromagnetic hair and differing by the
addition of soft photons. It has recently been suggested by Hawking~\cite{hawking}
that the apparent information loss paradox might be resolved by
considering supertranslations on the horizon, with the information `lost' by
incoming particles being recoverable {\it in principle}, though lost {\it in practice}.

This proposal raises many questions, including whether supertranslations
(together with superrotations and the corresponding electromagnetic gauge configurations) are capable
of encoding {\it all} the information carried by the incoming particles~\cite{dvali}, as well
as the relation to the stringy description of black holes, the details of the
mechanism for information transfer to and from the black hole, and whether
information is really lost.  We mention here for completeness
the recent work of `t Hooft~\cite{thooft}, where it is argued that the black-hole horizon
should be viewed as an apparent world-sheet of an induced string theory.
This is to be contrasted with the spirit of our approach in this paper, where we start from string theory at a fundamental level.
We also mention the recent work of Polchinski~\cite{polch} in which a shock-wave approximation
was used to calculate the shift on a generator of the horizon caused by an ingoing wave packet,
which is similar in spirit to our analysis below of the supertranslation of the horizon (viewed as a recoiling D-brane) induced by infalling matter. 
Other works in similar spirit include~\cite{englert,brus},  where the back-reaction of matter falling onto the black hole horizon (and fluctuations of the latter)
is argued to play an important  r\^ole in retaining information.

In this paper we review our previous arguments about the importance of
$W_\infty$ symmetry~\cite{iso,wcontr}~\footnote{This becomes $W_{1+\infty}$, if one includes conformal spin-one states.
We denote by lower-case $w_\infty$ the classical symmetry, and by upper-case $W_{1 + \infty}$ its quantum counterpart,
which includes an additional conformal spin-one state.} for `balancing the information books'. In particular,
we recall that a massless stringy particle interacting with the 
black hole is represented by a conformal operator on the world-sheet of the string,  which is exactly marginal 
{\it only if} a contribution from $W_\infty$ generators is included in its vertex function~\cite{emn1,Chlyk}.
Without this contribution, the corresponding renormalization-group (RG) $\beta$ function is non-zero,
leading to an inexorable increase in entropy. As we recall~\cite{emn1,emnQM}, $w_\infty$ (which is the classical limit of the quantum $W_\infty $ symmetry)
is the algebra of transformations that preserve the two-dimensional phase-space 
volume of massless (`tachyonic') stringy matter 
propagating in the background of a stringy black hole.  

Moreover, these quantum $W$-algebras are symmetries of the quantum scattering matrix of the corresponding two-dimensional string theory~\cite{klebanov,Chlyk}, in the sense that the operator product expansion between two appropriate vertex operators reproduce the corresponding $W$-algebra. In the flat space-time case (in which case the string theory is just a two-dimensional Liouville theory) the operators corresponding to the discrete higher-spin operators of the $W$ algebra are discretized `tachyon' operators. However, as already mentioned, in the presence of a black hole, at the quantum level,  the corresponding $W_\infty$ symmetries necessarily mix massless and 
massive (topological, delocalised) stringy states~\cite{Chlyk}. The admixture of $W_\infty$ generators
in the exactly marginal vertex operator of a massless string excitation shows how
information is transferred between a stringy black hole and external particles.

We discuss below the embedding of the two-dimensional coset describing the singularity in a four-dimensional space-time~\cite{emn2} with the structure SU(1,1)/U(1) $\otimes ~ S^{2}$,
where $S^2$ is a two-dimensional manifold with the topology of the sphere that is to be identified with the horizon of the four-dimensional black hole.
Under certain circumstances specified below,  in addition to the quantum $W$-symmetries that leave the (quantum-gravity) scattering matrix invariant, 
and are associated with the discrete (topological) states of the two-dimensional coset substructure, there is also a classical $w_\infty$ \ algebra of
symmetry transformations of the horizon coordinates that preserves the area of the horizon of an isolated spherically-symmetric four-dimensional
black hole, so that its entropy (which is known to be a Noether charge~\cite{wald}) is conserved.
The precise relation between these two different $W$-symmetries is not yet completely understood. It becomes evident, however, in the case where the horizon of the black hole
is represented as a thick (recoiling) brane, which is known to correspond to an SU($\infty$) gauge theory~\cite{fit,ilio}. The black hole emerging in the low-energy limit of this string/brane theory
is the \emph{infinitely-coloured} SU($\infty$) black hole discussed in \cite{winstanley}, which is reviewed below. In this case,
the gauge states on the horizon can be represented as open-string states whose ends are attached to the horizon brane, which then carry the SU($\infty$) charges. 
It is known that  classically such SU($\infty$) symmetries are isomporphic to the $w_\infty$ algebra that preserves a two-dimensional area, 
which can in this case can be identified with the horizon area of the spherically-symmetric, infinitely-coloured black hole. 

\section{Entropy Increase and Non-Critical String}

In the usual treatment of critical string theory, entropy remains constant and,
in particular, pure initial states remain pure during the time evolution. A stringy
discussion of the possibility of information loss therefore requires going beyond
the critical string framework. Accordingly, we now recall our arguments~\cite{emnQM} on
entropy increase in non-critical string theory, \emph{i.e.}, in a string model where conformal invariance 
on the world-sheet surface $\Sigma$ is broken by relevant operators. 

In such a case, the world-sheet 
dynamics depends on interaction terms that represent non-critical deformations of the form 
\begin{equation}\label{reldeform}
S_\sigma = S_\sigma ^\star + \int _\Sigma d^2 \xi \sqrt{-\gamma} \, g^i \, V_i 
\end{equation}
where $S_\sigma^\star$ is a conformal fixed-point $\sigma$-model action, summation over repeated indices is implied, $\gamma$ is a world-sheet metric, 
and the set of $\{ g^i \}$ is an (infinite in general) set of target-space fields associated with the 
corresponding vertex operators $V_i$. These target-space fields include the lowest-lying (massless) 
string modes, \emph{i.e.}, the graviton, dilaton and antisymmetric tensor fields, 
the scalar `tachyons' (in the case of a non-supersymmetric target space),
as well as the infinity of higher-spin states~\footnote{In a two target-space-dimensional setting, 
the only propagating  multiplet consists of massless scalar fields (misleadingly called `tachyons'),
whereas the graviton and higher-spin multiplets are topological states with discrete momenta. 
As we discuss later, such topological states exist also in higher-dimensional 
target space-times, so their presence is rather generic.}.

We introduce a global world-sheet RG scale ${\mathcal \mu}$,
and define ${\mathcal T} \equiv {\rm ln}\, \mu$. We consider a density matrix $\rho (g^i)$ describing the 
propagation of a string representing a matter state in the deformed target-space background corresponding to 
the world-sheet action (\ref{reldeform}). This density matrix is a generic function of the background fields $\{ g^i \}$. 
The renormalizability of the world-sheet two-dimensional theory implies that any explicit dependence of 
$\rho (g^i)$ on ${\mathcal T}$, represented by the corresponding partial derivatives,  is compensated 
by the `running' of the renormalised $g^i$ with ${\mathcal T}$, leading to the following world-sheet 
RG equation:
\begin{equation}\label{rge}
0 = \frac{d}{d {\mathcal T}} \, \rho (g^i ) = \frac{\partial}{\partial \, {\mathcal T}} \, \rho (g^i) + \frac{d\, g^j}{d\, {\mathcal T}} \, \frac{\partial \, \rho}{\partial \, g^j}~.
\end{equation}
It should be understood that the $\partial_i \equiv \frac{\partial}{\partial g^i} $ denote functional derivatives 
$\frac{\delta}{\delta g^i} $ with respect to the corresponding fields $g^i$ in target space-time. 
Roughly speaking,
\be\label{beta}
\frac{d\, g^j}{d\, {\mathcal T}}  \equiv \beta^i 
\ee
is a RG $\beta$ function for the `coupling' $g^i$ of the two-dimensional world-sheet field theory~\footnote{More 
strictly speaking, in string theory the target-space dependences of the `couplings' $g^i$ imply
some diffeomorphism variations, which lead to the replacement of the corresponding $\beta^i$
RG functions by the corresponding Weyl anomaly coefficients, but such complications are not 
relevant for our main arguments below, so we omit them here. For details see \cite{emn1}.}.

Representing (in a heuristic way) the density matrix by its Gibbs (equilibrium) form in target space, 
one has, in terms of the effective target-space string Hamiltonian 
\be\label{gibbs}
\rho (g^i) = {\rm Tr} e^{-\beta\, H}  \, ,
\ee
where $\beta$ is an effective `temperature' in target space, \emph{e.g.}, the Hawking temperature
in the case of a black hole. We assume
that the Tr operation commutes with $\partial/\partial \mathcal T$, which is consistent with the 
interpretation of the RG scale $\mathcal T$ as time (see below).
We then deduce that the von Neumann (entanglement) entropy, 
\be\label{entropy}
{\mathcal S} \equiv - {\rm Tr}\Big(\rho \, {\rm ln}\, \rho\Big)
\ee
varies with ${\mathcal T}$ as 
\be\label{dsdt}
\frac{\partial \, {\mathcal S}}{\partial {\mathcal T}} = - {\rm Tr} \Big[\frac{\partial \, \rho}{\partial {\mathcal T}} \, {\rm ln}\, \rho \Big]  - {\rm Tr} \Big[\frac{\partial \, \rho}{\partial {\mathcal T}} \Big] \, .
\ee
In \cite{emnQM} we have used Liouville dressing to restore conformal invariance on the world sheet,
and argued that the Liouville mode may be regarded as a local (on the world-sheet) covariant RG scale, 
$\rho(\sigma)$, where $\sigma$ denotes  the world-sheet coordinates.

In the presence of such a local world-sheet RG scale, there are counterterms in
the $\sigma$-model action of the form~\cite{emnQM}:
\be\label{ct}
\int_\Sigma\,  \partial_\alpha g^i  \, {\mathcal G}_{ij} \, \partial_\alpha g^j \, ,
\ee
where $\alpha = 1,2$ span the world-sheet coordinates. The only dependences on them
of the `renormalised' couplings $g^i$ occur through their dependences on the local RG scale $\rho (\sigma)$,
so we can write (\ref{ct}) as
\be\label{ct2}
\int_\Sigma \partial_\alpha \rho \, \partial_\alpha \rho \,   \hat \beta^i  \, {\mathcal G}_{ij} \,  \hat \beta^j \, ,
\ee
where $\hat \beta^i \equiv  d g^i/d \rho $ is a Weyl-anomaly coefficient. \emph{i.e.}, 
a world-sheet RG $\beta$-function with respect to the local RG scale. The quantity
${\mathcal G}^{ij}$ in (\ref{ct}, \ref{ct2}) acts as a Zamolodchikov `metric' in
the theory space $\{ g^j \}$ of the $\sigma$-model, \emph{i.e.}, in the space of target-space background fields. 
From conformal-field theory considerations~\cite{emnQM} that we do not discuss here, one has
\be\label{cft3}
\frac{d}{d\rho} Q^2[g] \propto \, \hat \beta^i  \, {\mathcal G}_{ij} \,  \hat \beta^j \, ,
\ee
where $Q^2 = C[g] - c^\star $
is the central-charge deficit of the corresponding non-critical string theory with central charge $C[g]$
that is a functional of $\{ g^i \}$, and the quantity $c^\star$ is the central charge at a conformal point. 

The local RG scale $\rho$ plays the r\^ole of a Liouville mode, which dresses the renormalised couplings 
$g^i$ in such a way so as to restore criticality in the 
$D+1$ dimensional target space: we find from (\ref{ct2}) that the scale $\rho$ is a propagating $\sigma$-model 
scalar field, and thus its zero mode on the world-sheet may be interpreted as an extra target-space-time coordinate.
The sign of $Q^2$ depends whether the theory is subcritical or supercritical, with the
supercritical case corresponding to $Q^2 > 0$. Actually, as discussed in detail in \cite{emnQM}, 
the dependence of $Q^2[g]$ on the local RG scale $\rho$ is such that the
derivative with respect to the world-sheet zero mode of $\rho$, $\rho_0$,  which is also identified with the global RG scale $\mathcal T = \rho_0$, obeys
$dQ^2/d\rho_0  \propto - Q^2 + \dots $, as a result of the fact that $Q^2$ is proportional to the 
$\sigma$-model partition function, in which the RG scale couples to the world-sheet curvature as 
$\int D\rho \, e^{- \int_\Sigma \rho \, R^{(2)} }$, where $\int_\Sigma R^{(2)}$ is the Euler characteristic 
of the world-sheet manifold. 
In this way, one recovers from (\ref{ct2}, \ref{cft3}) via perturbation theory the kinetic term of the Liouville action, 
\emph{i.e.}, a term in the $\sigma$-model-field-theory action $S_\sigma$ of the form
\be\label{ct4}
S_\sigma \ni -\int_\Sigma \partial_\alpha \rho \, \partial_\alpha \rho \,  ( Q^2 \,  \hat + \dots )
\ee
where, in our convention~\cite{emnQM}, the corresponding kinetic terms of the spatial target coordinates of the 
string $X^I$, $I=1, \dots 3$ have a minus sign. 

In this case the zero mode of the world-sheet RG flow of this super-critical string can be identified (up to a sign)
with the temporal flow in the target space-time~\cite{emnQM}:
\be\label{time}
t = - {\mathcal T} (= - \rho_0)~. 
\ee
With this identification, the second term on the right-hand side of (\ref{dsdt}) would
vanish in a theory with energy conservation on the average, as for quantum black holes. 

Then, using (\ref{rge}), (\ref{beta}) and (\ref{time}), we obtain from (\ref{dsdt}):
\bea\label{entropyscaling}
\frac{\partial \, {\mathcal S}}{\partial t} = - {\rm Tr} \Big[\frac{\partial \, \rho}{\partial \, t} \,  {\rm ln}\, \rho \Big] = {\rm Tr}\, \Big[
 \beta \, \frac{\partial H}{\partial \, t} \, \rho \, {\rm ln}\, \rho  \Big] =  - {\rm Tr} \Big[ \beta \, \beta^i \,  \frac{\partial H}{\partial g^i}\, \rho \, {\rm ln}\, \rho \Big]~,
\eea
where we have used (\ref{gibbs}) and the fact that $H (g^i)$ is a functional of the background fields. 

In a string theory setting, the effective Hamiltonian $H$ may be identified (up to a sign)
with the effective action $\Gamma$, and  it is known in this case that its field variations are
proportional off-shell to the $\beta^i$ (which is equivalent to the well-known statement that the 
string conformal invariance conditions on the world-sheet are equivalent on-shell to the 
equations of motion of the target-space effective action):
\be\label{zam}
\frac{\partial H}{\partial g^i} = - \frac{\partial \Gamma}{\partial g^i} = - {\mathcal G}_{ij} \, \beta^j ~,
\ee
where ${\mathcal G}_{ij}$ is the Zamolodchikov metric in the space of string models~\cite{emnQM}. 
From (\ref{entropyscaling}), (\ref{zam}) we then obtain
\be\label{increase}
\frac{\partial \, {\mathcal S}}{\partial \, t} =  \beta \beta^i \, {\mathcal G}_{ij} \, \beta^j \, \mathcal S~.
\ee
Within the context of a unitary world-sheet $\sigma$-model, corresponding to a
Euclidean target space-time as can be used to represent a finite-temperature black holes, 
the factor $\beta^i {\mathcal G}_{ij} \beta^j > 0$. In this case, then, (\ref{increase}) 
implies a \emph{monotonic} entropy increase of the (positive) entropy $\mathcal S > 0$ as time increases,
\emph{i.e.}, during the evolution from infrared to ultraviolet on the world sheet, whenever
a string propagates in a non-conformal background.

We use this result in the following, in the specific context of strings propagating in target-space black hole backgrounds.

\section{$W_\infty$ Symmetry Retains Information in a Two-Dimensional Stringy Black Hole}

The prototypical stringy black hole solution in two dimensions~\cite{witt} is characterized by
a world-sheet Wess-Zumino-Witten (WZW) $\sigma$-model formulated
on the coset space $SL(2,R)_k/U(1)$, where $k=9/4$ is the Kac-Moody algebra level.
The conformal invariance condition for this world-sheet $\sigma$-model induces a target-space 
metric corresponding to a Euclidean black hole background
\be\label{2bh}
ds^2 = dr^2 + {\rm tanh}^2 \, r \, d \tilde \theta^2 \, ,
\ee
where $(r, \theta)$ are two-dimensional coordinates, $r$ being the radial coordinate and
$\tilde \theta $ a compact `angular' coordinate that plays the r\^ole of an external temperature variable:
it should not be confused with a four-dimensional angular variable. 
The space time (\ref{2bh}) looks like a semi-infinite cigar, and may be elevated
to four dimensions via a similar formula with additional angular variables, as described in Section~\ref{sec:four}. 

The spectrum of stringy excitations in the two-dimensional stringy black hole includes
background massive topological states that possess a quantum $W_{1 +\infty}$ 
symmetry~\cite{iso,Chlyk}~\footnote{These symmetries were first discovered in operator product expansions of 
vertex operators corresponding to the discrete stringy `tachyon' states 
of the two-dimensional (target space) $c=1$ Liouville string theory~\cite{klebanov}
that, from a target space-time point of view, is the asymptotic limit of the Euclidean black hole of \cite{witt}.}.
This leads to an infinity of conserved charges (`hair') for the black hole system, rendering it completely integrable~\cite{witt}. 
The $W_{1 +\infty}$ symmetry corresponds classically to a classical infinite-dimensional $w_\infty$ 
algebra of diffeomorphisms that preserve a two-dimensional 
area form, which could correspond to the surface of a sphere $S^2$ or some other two-dimensional manifold.
In the case of the two-dimensional stringy black hole examined in this Section, this is
a symplectic phase space `area' form, which corresponds to the Hamiltonian of a particle system 
in the near-horizon geometry of the black hole. As we discuss later, this algebra appears also in the case
of four-dimensional stringy/brany solutions interpolating between black holes and AdS spaces.

Generically, a symplectic area two-form $\Omega$ corresponding to coordinates $x, y$:
\be\label{sympl}
\Omega = dy \wedge dx
\ee
is invariant under classical symmetry transformations that leave it
invariant. These area-preserving diffeomorphisms are generated by the quantities
\be\label{gen}
v_m^\ell = y^{\ell + 1} \, x ^{\ell + m + 1} 
\ee
where $\ell$ and $m$ are integers. The Poisson brackets of these generators 
satisfy the classical $w_\infty$ algebra
\be\label{winfty}
\{ v_m^\ell, \, v_{m^\prime}^{\ell^\prime} \} = [m \, (\ell^\prime + 1) - m^\prime \, ( \ell + 1) ]\, v_{m + m^\prime}^{\ell + \ell^\prime} \, .
\ee
This includes a Virasoro symmetry generated by the operators $L_n = v_{n}^{0}$,
whose Poisson brackets obey the algebra
\be
\{ L_n, \, L_m \} =  (m-n)\, L_{m+n}~,
\label{Vira}
\ee
which is a subalgebra of the $w_\infty$ algebra (\ref{winfty}).

We suggested in \cite{emn1} that the infinite set of charges appearing in the quantum version of the $w_\infty$ symmetry
should be considered as an infinite set of discrete hair (termed W-hair) that is responsible for the maintenance of quantum coherence for the two-dimensional stringy black hole, since
the corresponding quantum-gravity scattering matrix, 
obtained from correlation functions of marginal world-sheet vertex operators, is invariant under these symmetries.

Consider, for example, the propagation of a `tachyon', which is a massless particle
in two dimensions. In flat space it is associated with the vertex operator:
\be
             \phi^{c,-c}_{-1/2,0,0} =(g_{++}g_{--})^{-\frac{1}{2}}
F(\frac{1}{2} ; \frac{1}{2} ; 1 ; \frac{g_{+-}g_{-+}}{g_{++}g_{--}}) ,
\label{XV}
\ee
where $F$ denotes a hypergeometric function and $g_{ab}$, $a,b =+,-$ represent the components
of a generic $SL(2,R)$ element. The operator (\ref{XV}) is exactly marginal in a two-dimensional flat-space string theory.

However, this is not the case in the background of a two-dimensional space-time black hole 
(Euclidean or Minkowski, the latter being obtained by analytic continuation of the compact variable (`temperature') 
in the cigar metric of \cite{witt}).
In this case, the corresponding  \emph{exactly marginal} operator is~\cite{Chlyk}
\be
          L^1_0 {\overline L_0}^1 = \phi^{c,-c}_{-1/2,0,0}+ i (\psi ^{++}- \psi^{--}) + \dots
\label{XVI}
\ee
where
\be
  \psi^{\pm\pm}   \equiv : ({\overline J}^{\pm} )^N (J^{\pm})^N  (g_{\pm\pm})^{j+m-N} :
\label{XVII}
\ee
with
$J^{\pm} \equiv (k-2) (g_{\pm\mp}\partial _z g_{\pm\pm} -
g_{\pm\pm}\partial _z g_{\pm\mp} )$, and
${\overline J}^{\pm} \equiv (k-2) (g_{\mp\pm}\partial _{{\bar z}}
g_{\pm\pm} - g_{\pm\pm}\partial _{{\bar z}} g_{\mp\pm})$,
where $k$ is the WZ model level parameter \cite{witt}.
The combination $\psi^{++} - \psi^{--}$ generates a level-one
massive string mode, and the dots in equation (\ref{XVI})
represent operators that generate higher-level massive string states~\footnote{Another
example of an exactly-marginal operator is $L^2_0 {\overline L_0}^2   = \psi^{++} + \psi^{--} + \psi^{-+}
+\psi^{+-} + \dots$, which also involves in an essential way operators for massive string
modes. The coupling corresponding to this world-sheet deformation of the coset model
is associated with a global rescaling of the target space-metric \cite{Chlyk}, and therefore to a global
constant shift of the dilaton field. Thus it produces shifts in the black hole mass~\cite{witt}.}.
As discussed in~\cite{emn1}, these modes are solitonic, with fixed energy and momentum. As such, they are
completely de-localized in space-time.

Since the flat-space `tachyon' vertex operator (\ref{XV}) is not exactly marginal in a black-hole
background, the corresponding RG $\beta$ function is non-vanishing and hence, by (\ref{increase}),
the entropy associated with tachyonic `matter' increases inexorably, \emph{i.e}, information is lost, \emph{if the
higher-level string modes in (\ref{XVI}) are neglected.} Conversely, if these string modes are
taken into account, the corresponding RG $\beta$ function vanishes, entropy does not increase
with the world-sheet RG flow, which we identify with the target-space temporal time flow in our approach.
Thus, there is no information loss: it is stored by the higher-level string modes.

In order to guarantee the exact marginality of the corresponding vertex operator (\ref{XV}),
topological states must be included in the scattering matrix of strings in a two-dimensional
black-hole background. These topological modes are not detectable in a local scattering experiment, 
leading to an apparent `loss' of quantum coherence, which is an artefact of the phenomenological
truncation of the scattering process within a local effective field theory (LEFT) framework. 
Associated with this apparent `loss' of quantum coherence there is an apparent `increase'
in entropy at a rate quantified by the right-hand-side of (\ref{increase}), since the \emph{truncated} RG 
$\beta^i$ functions of the non-marginal propagating modes do not vanish. 

Nevertheless, the conserved W-hair charges are in principle measurable, 
and ways for doing so in principle have been outlined in \cite{emn2}. These are reviewed in the next Section, where we
also present arguments for the elevation of the W-hair to four-dimensional space times.
In this case, the area-preserving property of the $W_\infty$ symmetry becomes important for preserving the area of 
the two-dimensional surface of the black-hole horizon.

\section{Elevation to Four Dimensions \label{sec:four}}

We have argued in~\cite{emn2} that the coset singularity structure of the two-dimensional stringy black hole and
generic properties of its associated discrete states have counterparts
for spherically-symmetric black-hole configurations in four space-time dimensions.
We now review the basic arguments supporting this conjecture, which have been reinforced by subsequent
formal developments.

\subsection{Embedding of the Two-Dimensional Black Hole}

We consider  a string theory with a spherically-symmetric gravitational
background of black-hole type, which is a solution of the Einstein equations, generalised to the effective
field theory derived from string theory. The metric tensor is given by an Ansatz of the form:
\begin{equation}
ds^2 =g_{\alpha\beta} dx^{\alpha}dx^{\beta} + e^{W(r,t)}d\Omega^2 \, ,
\label{sphere}
\end{equation}
where W(r,t) is a non-singular function, $x^{\alpha,\beta}$ denote the $r,t$ coordinates,
and $d\Omega ^2 = d\theta ^2 + \sin^2 \theta d \phi ^2 $ denotes the line element
on a spherical surface that does not change with time.

We remind the reader that in pure gravity all the {\it classical} spherically-symmetric solutions to the equations of
motion obtained from higher-derivative gravitational actions with an arbitrary number of curvature tensors
are {\it static}~\cite{whitt}, and that a similar result holds for stringy black holes at tree level.
The standard Schwarzschild solution of the spherically-symmetric four-dimensional
black hole can be put in the form (\ref{sphere}) by an appropriate transformation of variables.

We consider the Schwarzschild solution in Kruskal-Szekeres coordinates~\cite{thorn}
\begin{equation}
ds^2 = -\frac{32M^3}{r}e^{-\frac{r}{2M}}du dv + r^2 d\Omega ^2 \, ,
\label{krusk}
\end{equation}
where $r$ is a function of $u,v,$ given by
\begin{equation}
  (\frac{r}{2M} - 1) e^{\frac{r}{2M}}=-uv \, .
\label{reln}
\end{equation}
Note that, although the two-dimensional metric components depend
on the variables $u,v$, the black hole solution is nevertheless static.
Changing variables to
\begin{eqnarray}
\nonumber
e^{-\frac{r}{4M}}u=u' \, , \\
e^{-\frac{r}{4M}}v=v'
\label{change}
\end{eqnarray}
and
taking into account the Jacobian $J$ of the transformation of the area element $dudv$, we
can put the two-dimensional metric in the form
\begin{equation}
 g_{bh}(u',v')= \frac{e^{D(u',v')} du'dv'}{1-u'v'} \, ,
\label{bh}
\end{equation}
with the scale factor being given by $ 16M^2 e^{-\frac{r'(u',v')}{2M}}J(u',v')$,
where $r'$ is the coordinate $r$ re-expressed in therms of the coordinates $u',v'$.

The metric (\ref{bh}) is a conformally-rescaled form of Witten's two-dimensional  black hole solution~\cite{witt}. Since the
latter is described by an exact conformal field theory, the same is true after this conformal rescaling.
From a $\sigma$-model point of view, this rescaling simply expresses a change of renormalisation scheme~\footnote{The 
function $D(u,v)$ can be regarded also as a part of the two-dimensional dilaton in the given renormalisation
scheme.}.  The global properties, such as singularities, remain unchanged
from the two-dimensional string case.

\subsection{Discrete Topological States}

In particular, the infinite-dimensional W-symmetry associated with the
SU(1,1)/U(1) coset structure of the dilaton-graviton sector in the two-dimensional model
is elevated to become a model-independent feature of spherically-symmetric
four-dimensional string configurations. Such structures are intimately connected with 
the existence of topological solitonic non-propagating states. 
These states are essentially 
spherically-symmetric solutions of the low-energy equations of motion obtained from the string theory in manifolds 
with topology SU(1,1)/U(1) $\times {\mathcal M}^2$,
where ${\mathcal M}^2$ is a two-dimensional manifold of constant curvature.
They are associated with jumps in the number of degrees of freedom at discrete values of energy and 
momentum as a result of relaxation of certain gauge theory constraints,
as shown below. 
The simplest example is where
${\mathcal M}^2=S^2$, the sphere, which describes the spherically-symmetric four-dimensional black hole
solution of interest to us here.  
The associated infinity of discrete topological (non-propagating) states, with definite energies and momenta,
couple to the massless propagating `tachyon' string matter and thereby ensure conformal invariance of the 
associated $\sigma$-model action, as described above for the purely two-dimensional stringy black hole 
of~\cite{witt}.  

The infinity of discrete topological states in a  $D$-dimensional target-space
string theory are similar in nature to those of the two-dimensional case~\cite{pol,klebanov}. These states can be seen via
the gauge conditions for a rank-$n$ tensor multiplet:
\begin{equation}
   D^{\mu_{1}}A_{\mu_{1}\mu_2...\mu_n}=0 \, ,
\label{gauge}
\end{equation}
where $D_{\mu}$ is a (curved-space) covariant derivative.
To illustrate our arguments,
consider the simplified case of
weak gravitational perturbations around
flat space, with a linear dilaton field
of the form $\Phi(X)=Q_{\mu}X^{\mu}$, in which case the
Fourier transform of (\ref{gauge}) is
\begin{equation}
    (p + Q)^{\mu_1} {\tilde A}(k)_{\mu_1\mu_2....\mu_n}=0
\label{gauge2}
\end{equation}
We observe that there is a jump in the number
of degrees of freedom at the discrete momentum $p=-Q$.
The fixed momentum corresponds to
complete uncertainty in space, so such states are delocalised,
and can be considered as quasi-topological and non-propagating
soliton-like states. In ordinary string theories, such states
carry a small statistical weight, due to the continuous
spectrum of the various string modes. However,
when strings propagate in spherically-symmetric four-dimensional
background space-times, these discrete states
assume particular importance. Such backgrounds are effectively
two-dimensional, and therefore all the
transverse modes of higher-rank tensors can be gauged away
using Ward identities of the form (\ref{gauge}), except for
the {\it topological} modes. In the case of four-dimensional
spherically-symmetric black holes, these $s$-wave
topological modes constitute the final stages
of their evaporation \cite{emn1}, and assume
responsibility for the maintenance of
quantum coherence \cite{emn1,emn2}. 

\subsection{Phase-Space-Area-Preserving $w_\infty$ Symmetries}

In another example~\cite{zanon}, a $w_\infty$ symmetry arises in the phase space of matter in a 
four-dimensional extremal solitonic black hole background in the context of 
$N=2$, $D=4$ supergravity.
This is a BPS solution that interpolates between a maximally-supersymmetric AdS$_4$
space-time at large radial distances and AdS$_2 \times H^2$, where AdS$_2$ refers to the radial-coordinate/time
part of the space-time and $H^2$ refers to the angular part of the space-time, 
which is a hyperbolic two-dimensional manifold of constant curvature. The AdS$_2 \times H^2$
geometry characterises the space-time near the horizon of the black hole. The analysis of \cite{zanon} 
showed that the dynamics of a quantum-mechanical  massive particle with non-trivial magnetic charge
in the near-horizon geometry is described effectively by a one-spatial-dimensional Hamiltonian $H$, 
characterised by a $w_\infty$ symmetry that preserves the two-dimensional phase-space area symplectic form
$\Omega  = dp \wedge dq - dH \wedge dt $, with $q$ the spatial coordinate, $p$ the canonical momentum and $t$ the time. 
The energy spectrum of this particle is continuous and bounded from below: $E>0$,
but  the ground state is non-normalizable, with an infrared (IR) divergence,
which was regularised in \cite{zanon} by putting the system in a box.
The IR-regularised system is also invariant under a $w_\infty$ that contains a Virasoro symmetry (\ref{Vira}),
which can be associated 
with the asymptotic symmetries of the AdS$_2$ space time,
{i.e.}, the diffeomorphisms that leave invariant the AdS$_2$ metric, whose
quantum version includes a central extension. 
Such asymptotic symmetries are symmetries of the quantum-gravity scattering matrix for the full 
four-dimensional AdS$_2 \times H^2$ extremal black hole of \cite{zanon}~\footnote{An asymptotic 
symmetry of the quantum-gravity scattering matrix under supertranslations of generic black hole 
backgrounds has been examined in \cite{strominger,strominger2}.}.

Hence, the particle system is characterised by an infinity of conserved charges of the $v_m^\ell$ type (\ref{gen}), 
in which the r\^ole of the $x,y$ coordinates is played by appropriate combinations of the phase-space 
coordinates of the particle~\cite{zanon}, and hence is completely integrable.
From our point of view, the presence of an infinity of conserved quantities for the 
particle in the near-horizon geometry of the black hole also guarantees quantum coherence, 
in the sense that the infinity of conserved charges $v_m^\ell$, 
which remain constant during the scattering of matter off the black-hole background,
retain information during the evaporation of the latter.  
The situation of the coherence-preserving $w_\infty$ algebra is exactly analogous to that 
preserving the phase-space area for a massless `tachyonic' string matter 
in the two-dimensional stringy black hole - or its four-dimensional extension with topology 
SU(1,1)/U(1) $\times S^2$ - as discussed above. 

As discussed above, the elevation of such classical phase-space-area- 
$w_\infty$ symmetry algebras 
to fully quantum coherence-preserving algebras necessarily involves discrete topological states of the string. 
In two dimensions, as we have seen, the latter mix with the propagating massless matter
states in order to guarantee the conformal invariance of the corresponding vertex operators in the presence of a stringy black hole background~\cite{emn1}, and hence preserve quantum coherence according to the general arguments
of Section~2. 

\subsection{W-hair and Quantum Coherence}

As discussed in  \cite{wmeasure}, each
of these discrete solitonic states can be represented
as a singular gauge configuration,
whose conserved $W$-charges can be
measured in principle by generalized Aharonov-Bohm phase
effects. Moreover, the topological higher-spin string states leave their imprint via
selection rules in the scattering matrix, where they appear as (resonance) poles, 
corresponding to discrete energies and momenta and leading to certain selection rules. 
In the stringy black hole case, there is an infinite
set of such black hole soliton states, classified by the quadratic
Casimir and `magnetic' quantum numbers of an internal symmetry
group~\cite{wmeasure}, which are excited at calculable energies and decay into
distinctive combinations of light final-state particles. 

The stringy scattering matrix is, in general, well defined in the presence of such black-hole backgrounds, 
since the world-sheet correlation functions among the appropriate exactly marginal vertex operators are unitary. 
This is because, as mentioned previously, in addition to the parts corresponding to the propagating string states,
these operators contain an infinity of topological non-propagating states. In practice, 
scattering experiments in the laboratory, which involve a finite number of localised (in spacetime) particle states, 
cannot detect the delocalised states.
Hence, from the point of view of a local low-energy observer, there would be an apparent decoherence, 
although this would not entail any pathologies in the full stringy theory of quantum gravity.

\section{Phase-Space vs Horizon-Area-Preserving $W$ Symmetries}

We now explore the potential relation between the quantum $W_{1+\infty}$ algebras 
that are symmetries of the stringy quantum gravity S-matrix 
and the classical area-preserving symmetries that preserve the horizon area of a 
classical (non-evaporating) black hole and hence its entropy. 
This relation is subtle, and at present is not understood  in its full generality, at least by the authors. 
Nevertheless, as we shall discuss below, 
by representing the horizon of the black hole as a thick D(irichlet) brane, 
such a relation becomes evident. As a prelude to this result, we first discuss the case of an
infinitely-coloured four-dimensional black hole in a SU($N \to \infty$) Yang-Mills gauge theory, which has an 
infinite amount of gauge hair, as allowed by the `no-hair' theorem~\cite{winstanley}.

\subsection{Black Holes with Infinitely-Coloured Hair \label{sec:suinfty}} 

As we have discussed, the classical $w_\infty$ algebra preserves the two-dimensional area of an `\emph{internal space}' with the topology of a 
sphere~\cite{iso,wcontr}. The issue is whether the `\emph{internal}' sphere can be identified with the 
real horizon of the spherically-symmetric four-dimensional Schwarzschild black hole.
To address this question, we consider examples of four-dimensional spherically-symmetric black holes with infinitely-coloured hair, which 
realize explicitly a classical $w_\infty$.  
These appear in an effective field theory example of a black-hole solution in SU($N\to \infty$) gauge theory in a 
four-dimensional AdS space-time with negative cosmological constant, which plays the role of a regulator for the black-hole solution
that makes it well-defined~\cite{winstanley}. 
This anti-de-Sitter (AdS) regulator was given physical significance via the AdS/CFT bulk/boundary correspondence, 
and turns out to be physically important, as we argue below. 
For the present discussion, the interest of these black holes with black holes with infinitely coloured hair is that
classically there is an isomorphism between SU($N \to \infty$) and $w_\infty$~\cite{iso,fit}~\footnote{For SU(N) gauge theories with finite $N$,
the geometry of the corresponding space is non-commutative~\cite{ilio},
the commutativity being restored in the limit $N \to \infty$.}. 

To develop this point, consider a unit sphere $S^2$ with coordinates $\theta, \phi$ and the quantities:
\be\label{xs}
x_1 = {\rm sin}\theta \, {\rm cos}\phi, \quad x_2 =  {\rm sin}\theta \, {\rm sin}\phi, \quad x_3 = {\rm cos}\theta~, \quad {\rm with} \quad 
\sum_{i=1}^3 x_i^2 = 1~.
\ee
The spherical harmonics $Y_{\ell\, m}(\theta, \phi)$ are harmonic polynomials of degree $\ell$ in $x_i$:
\be\label{harm}
Y_{\ell\, m}(\theta, \phi) = {\sum_{i_k=1,2,3}}_{k = 1, \dots ,\ell}\, \alpha^{(m)}_{i_1 \dots i_\ell} \, x_{i_1} \dots x_{i_\ell}~.
\ee
For fixed $\ell$ there are $2\ell + 1 $ linearly independent symmetric and traceless tensors
$\alpha^{(m)}_{i_1 \dots i_\ell}$, $m=-\ell, \dots , \ell$. Let us consider an SU(2) subgroup of the SU(N) group
in the limit $N \to \infty$, generated by 
$S_i$ with standard commutation relations
$$ \Big[ S_i\, , \,S_j \Big] = i \epsilon_{ijk}\, S_k~. $$
From the standard theory of angular momentum~\cite{schwinger,ilio}, we know that a representation of the $N^2 -1$ generators of the group SU(N)
can be expressed as follows in terms of the $S_i$ matrices and the $\alpha$-tensors in (\ref{harm}):
\begin{eqnarray}\label{suN}
S_{\ell \, ,\, m}^{(N)} &=& {\sum_{i_k =1,2,3}}_{k=1, \dots, \ell} \alpha_{i_1 \dots i_\ell}^{(m)} \, S_{i_1} \dots S_{i_\ell}  \nonumber \\
 \Big[ S_{\ell, \, m}^{(N)}\, , \, S_{\ell^\prime, \, m^\prime}^{(N)}  \Big]  &=&
 i f_{\ell, \, m;\, \ell^\prime, \, m^\prime}^{(N)\, \ell^{\prime\prime}, \, m^{\prime\prime}}\, S_{\ell^{\prime\prime}, \, m^{\prime\prime}}^{(N)}~.
\end{eqnarray}
Upon the rescaling
\be\label{rescale}
S_i \, \rightarrow \,  T_i \equiv \frac{2}{N}\, S_i ~,
\ee
we arrive at bounded matrix elements as $N \to \infty$: $|(T_i)^a_{\,\, b}| \le 1$, with the well-defined commutator algebra 
\be\label{tcomm}
\Big[ T_i\, , \, T_j \Big] = i \frac{2}{N}\, \epsilon_{ijk}\, T_k \, \rightarrow 0, \, N \to \infty~.
\ee
and the Casimir element 
\be\label{casimir}
T^2 = \sum_{i=1}^{3} T_i^2 = 1 - \frac{1}{N} \, \to \, 1 , \, N \to \infty~. 
\ee
We conclude from (\ref{tcomm}) and (\ref{casimir}) that in the $N \to \infty$ limit, a representation of the (commuting) generators of the $SU(2)$ subgroup of $SU(\infty)$ is provided by the quantities $x_i$ in (\ref{xs}). If one considers any two functions of $x_i$ on a spherical surface, 
$f(x_1,  x_2, x_3), g(x_1, x_2, x_3)$, each of which can be expanded in terms of the spherical harmonics (\ref{harm}), (\ref{tcomm})
shows that in the limit $N \to \infty$ the corresponding matrix polynomials of the generators $f(T_1, T_2, T_3)$ and $g(T_1, T_2, T_3)$ satisfy
\be\label{poisson}
\frac{N}{2i} \, \Big[ f, \, g \Big]  \, \rightarrow  \, \epsilon_{ijk}\, \frac{\partial \, f}{\partial \, x_j} \, \frac{\partial g}{\partial \, x_k}~, \quad N \to \infty \, .
\ee
Replacing the SU(2) generators in (\ref{suN}) by the rescaled ones (\ref{rescale}),
one finds for the $N^2 -1$ matrices $T_{\ell\, m}^{(N)}$:
\be\label{compoisson} 
\frac{N}{2i} \, \Big[ T_{\ell, \, m}^{(N)}\, , \, T_{\ell^\prime, \, m^\prime}^{(N)}  \Big]  \rightarrow \,  \quad \{ Y_{\ell, \, m} \, , \, Y_{\ell^\prime, \, m^\prime} \}, \quad N \to \infty \, ~.
\ee
The Poisson (classical) algebra of the spherical harmonics is known to be that of
the infinite-dimensional area-preserving diffeomorphisms  on the sphere $SDiff(S^2)$:
\bea\label{sdiffs}
\{ Y_{\ell, \, m} \, , \, Y_{\ell^\prime, \, m^\prime} \} &=&\frac{M(\ell + \ell^\prime -1, \, m + m^\prime)}{M(\ell, \, m)\, M(\ell^\prime, \, m^\prime)}\, (\ell^\prime \, m - \ell \, m^\prime) \, Y_{\ell + \ell^\prime -1, \, m + m^\prime}  \nonumber \\
&+ & \sum_{n=1} \, g_{2n}(\ell, \, \ell^\prime)\, C^{\ell + \ell^\prime -1 -2n, \, m + m^\prime}_{\ell, \, m, \, \ell^\prime, \, m^\prime} \, Y_{\ell + \ell^\prime -1-2n, \, m + m^\prime}~,
\eea
where the $M $, $g_{2n}$  are normalization functions and the structure constants $C$ are given in the fourth paper in~\cite{iso}. 
This algebra  is known to be isomorphic to the classical area-preserving $w_\infty$ algebra.

Since the classical gauge fields of the SU($N \to \infty$) gauge theory can be expanded in the basis of the matrices $T_{\ell, \, m}^{(N)}$, 
the above considerations, and in particular (\ref{compoisson}), indicate that in this example of an infinitely-coloured gauge black hole,
this area-preserving diffeomorphism symmetry preserves the horizon area, once we identify the `internal' sphere $S^2$ with the 
actual horizon sphere of the spherically-symmetric SU($\infty$) black hole. In this case, the entropy of the black hole
can be preserved \emph{classically} by the $w_\infty$ hair. If one views this SU($\infty$) gauge theory as 
a low-energy limit of some string theory then, in view of our world-sheet renormalization-group interpretation of the target time
that leads to (\ref{increase}), the conservation of the classical area should correspond to the conformal invariance of the corresponding world-sheet, 
which guarantees the vanishing of the right-hand-side of (\ref{increase}) through the zeroes of the $\beta^i$ functions of appropriate combinations 
of the couplings $g^i$. In the case at hand, the set $\{ g^i \}$ consists of the graviton $G_{\mu\nu}$ and SU($\infty$) gauge field background
modes: $A_\mu^a, a=1 \dots \infty$.

Hence one can understand the entanglement of the massless (propagating) graviton states in this four-dimensional extremal black hole
with the discrete infinity of gauge states by analogy with the entanglement of the propagating massless `tachyonic' matter 
with the infinity of the discrete massive states in the two-dimensional black hole case,  (\ref{XV}, \ref{XVI}), as follows.

Upon embedding the SU($\infty$) gauge theory in a string model with propagating graviton and gauge field backgrounds, 
one deforms the corresponding $\sigma$-model by adding  to the usual graviton (spin-2) deformations the following 
vector deformation of the gauge field  $A_\mu^a$:
\be\label{ gauge}
\mathcal Z = \int [DX] \, e^{\frac{1}{2\pi \alpha^\prime} \int_\Sigma \sqrt{\gamma} \partial_\alpha X^\mu \, \partial^\alpha X^\nu \, G_{\mu\nu}(X)} \, {\rm Tr} \Big(e^{\int_{\partial \Sigma} A_\mu^a t^a \partial_\tau X^\mu } \Big)~.
\ee
Here we use the standard notation for a $\sigma$-model propagating on a world-sheet $\Sigma$ with a boundary $\partial \Sigma$ 
(to accommodate open strings corresponding to gauge field excitations), describing the motion of a string in a target-space with 
coordinates $X^\mu$ and a metric background  $G_{\mu\nu}(X)$. The trace Tr is over colour indices, 
and the $t^a$ are the generators of the SU($\infty$) colour group.

The presence of an AdS background with a non-vanishing cosmological constant $\Lambda < 0$ as 
a regulator for the black hole of \cite{winstanley} implies that, on embedding such a theory in a string theory, 
the graviton world-sheet $\beta$-function, $\beta_{\mu\nu}^G$,  is no longer zero. Indeed, a $\sigma$-model one-loop analysis, 
which suffices for the weak gravity in the near-horizon black-hole geometry that we consider here, is given by the target-space Ricci tensor:
\be\label{gbf}
\beta_{\mu\nu}^G = R_{\mu\nu} \, ,
\ee
which for an anti de Sitter space time reads
$$ R_{\mu\nu}^G = \Lambda g_{\mu\nu} \ne 0~, \Lambda < 0~.$$
If this divergence were not cancelled against other background fields, so as to restore marginality of the corresponding
world-sheet operator, then there would be entropy increase under the RG flow, which in our approach is identified with real-time flow (\ref{increase}), 
and thus information loss in the classical stationary black-hole background. 

The presence of an infinity of gauge fields can provide a resolution to this problem 
with the infinity of zero-momentum gauge field modes $A_\mu^a, a=1, \dots N \to \infty$ condensing: 
\be\label{qcd}
\langle \sum_{a=1}^\infty \, F_{\mu\nu}^a F^{a \mu\nu} \rangle \ne 0 \, .
\ee
The infinite number of colours plays a crucial r\^ole in guaranteeing a macroscopic occupation of the quantum-mechanical ground state of this system, 
which is a prerequisite for the formation of a quantum condensate. One needs delocalised zero modes, 
because they are constant in space-time, and hence their condensation guarantees the space-time translational invariance of the condensate. 
In string theory there is an infinity of higher-order self-interactions among the gauge fields in the low energy string effective action, 
which can lead to the formation of such a condensate. 

Coupling gauge fields and gravitons, the corresponding graviton $\beta$-function (\ref{gbf}) 
is modified in the presence of such a condensate (\ref{qcd}) to
\be\label{cgbf}
\beta_{\mu\nu}^G = \Lambda \, g_{\mu\nu} + \frac{1}{2} V(\langle F_{\mu\nu}^a \, F_{\mu\nu}^a \rangle ) \, g_{\mu\nu} \, ,
\ee
where $V(\langle F_{\mu\nu}^a \, F_{\mu\nu}^a \rangle )$ is the scalar vacuum energy arising from the condensate. 
The structure on the right-hand side of (\ref{cgbf}) is the only one consistent with Lorentz invariance of the vacuum.  
The form of $V(\langle F_{\mu\nu}^a \, F_{\mu\nu}^a \rangle )$ is, to lowest order in the field strengths:
\be\label{vexpl}
V(\langle F_{\mu\nu}^a \, F_{\mu\nu}^a \rangle) \propto \langle F_{\mu\nu}^a \, F_{\mu\nu}^a \rangle + \dots > 0  \, ,
\ee
where the $\dots $ indicate higher derivative terms that are present in string theory and are essential in providing the 
necessary self-interactions among the non-Abelian gauge fields to guarantee the formation of condensates. 

We observe from (\ref{vexpl}) and (\ref{cgbf}) that a cancellation of the right-hand-side is possible because $\Lambda < 0$ for AdS backgrounds.
Thus we recover a conformal graviton background in the presence of a condensate formed by the delocalised gauge states.
The latter realize a $W_\infty$ algebra, which preserves classically the horizon area of the black hole, 
thus providing a picture of the area-preserving nature of the W-hair that is consistent
with the conformally-invariant mixing of graviton states with the infinity of gauge states. 

\subsection{Horizons as `Thick' D(irichlet)-Branes and SU($\infty$) Gauge Theory}

If we represent the horizon of the four-dimensional black hole as a two-brane, then we immediately face the problem of recoil once a string matter state, 
represented by a closed or open string, encounters the horizon surface. If it is a closed string, it may split into two open strings (to preserve the chirality of the state), whereas if it is an open 
string then at least one of its ends will be attached to the horizon, causing the latter to recoil in order to conserve momentum. 
The recoil of the horizon induces local fluctuations on the horizon that can be studied using logarithmic conformal field theory on the world-sheet~\cite{kogan, EllisDbrane}, 
that have also been argued to carry information~\cite{mislays}. A fluctuating (recoiling) horizon may be represented (from the point of view
of a low-energy observer) as a `thick' D-brane stack of $N \to \infty$ concentric branes. For macroscopic black holes, with large horizons compared to the wavelength of the infalling matter, 
such concentric branes may be well approximated locally by a stack of parallel flat $N \to \infty$ branes. Such constructions are equivalent to SU($N \to \infty)$ gauge theories~\cite{maldacena}, 
as can be seen intuitively by considering the topologically-equivalent ways ($N^2-1$ for SU(N) gauge theory) in which an open string can be attached to a stack of $N$ parallel D-branes.  

When infalling matter crosses the horizon of such a thick-horizon-brane black hole, the recoil is described by open string excitations that
carry the SU($\infty$) charges, leading to the infinite hair of the black hole and corresponding to the horizon-area preserving $w_\infty$ symmetry
discussed previously in Subsection \ref{sec:suinfty}.
The important aspect of this example is that now the SU($\infty$) symmetry is also a coherence-preserving symmetry of the associated 
quantum-gravity scattering matrix in the presence of the SU($\infty$) black hole. 

\subsection{Two-Dimensional $W_\infty$ Symmetries as Gauged Four-Dimensional Symmetries \label{sec:wgauge}}

In this subsection we speculate on a potential generalization of the above result to arbitrary four-dimensional space-times with space-time singularities having the (stringy-black-hole-like) structure 
SU(1,1)/U(1) $\otimes ~ S^{2}$ or, more generally, embedding two-dimensional singularity structures admitting $W_\infty$ symmetries into spherically-symmetric 
four-dimensional space-times with constant curvature, as in the example of the four-dimensional black-hole soliton of \cite{zanon}. 

Our starting point is the construction~\cite{wcontr} of $W_\infty$ (and $w_\infty$) gauge theories in terms of $(d+2)$-dimensional local fields, 
where $d$ is the dimension of space-time on which the algebras live: $d=2$ in the case of interest to us. One can define the $W_\infty$ quantum algebra as a 
commutator algebra of Hermitian operators $\xi (a, a^\dagger)$, where $a, a^\dagger$ are the harmonic-oscillator annihilation and creation operators. 
One may parametrize the operators $\xi(a, a^\dagger)$ using coherent states:
$$ :\xi (\hat a, {\hat a}^\dagger): = \int d^2z \, e^{-|z|^2} \, |z> \, \xi (z, \overline z) \, <z|~,  $$
where $|z> = e^{{\hat a}^\dagger \, z} \, |0>, \, <z| = <0| \, e^{\hat a \, \overline z}, \, <z^\prime | z > = e^{{\overline z}^\prime\, z}, \, \hat a\, |z> = z \, |z>,  \, <z| \, {\hat a}^\dagger = < z|\, \overline z$,
and the normalization condition is $\int d^2 z\,e^{-|z|^2} |z>\,<\overline z| = 1,$ with $d^2 z \equiv  \frac{1}{\pi} {\rm Re}\, z \, {\rm Im} \, z$, 
and the $:\xi(\hat a, {\hat a}^\dagger):$  is an (anti-)normal-ordered operator, where the creation operators are always placed to the right of the annihilation operators. 

The coordinates $z, \overline z$ are viewed in \cite{wcontr} as a group-theoretical (`colour') space. Introducing a gauge potential $A_\mu (x, \hat a, {\hat a}^\dagger) $, 
where $\mu = 1 , \dots d$ is a $d$-dimensonal space time $\{ x \}$ index,
\be\label{gp}
\hat A_\mu (x) \equiv A_\mu (x, \hat a, {\hat a}^\dagger) = \int d^2 z \, e^{|z|^2} \, |z> \, A_\mu (x, z, \overline z) \, < z|
\ee
and infinite-dimensional set of infinitesimal $W_\infty$ gauge transformations
can be introduced as follows:
\be\label{gtr}
\delta \hat A_\mu (x) = \partial_\mu \hat \xi (x) + i \Big[\hat \xi (x), \, \hat A_\mu(x) \Big]~, \quad
\delta A_\mu (x, z, \overline z) = \partial_\mu \xi (x, z, \overline z) - \{\{ \xi, \, A_\mu \}\}_{\rm Moyal}(x, z, \overline z)~,
\ee
where $\{\{ ., . \}\}_{\rm Moyal}$ denotes a Moyal bracket, defined as 
\be\label{moyal}
\{\{ \xi_1, \, \xi_2 \}\}_{\rm Moyal}(z, \overline z)  \equiv i \sum_{n=1}^{\infty} \frac{(-1)^n}{n\!} \Big( \partial_z^n \xi_1 (z, \overline z) \, 
\partial_{\overline z}^n \, \xi_2 (z, \overline z) - \partial_{\overline z}^n \xi_1 (z, \overline z) \, 
\partial_{z}^n \, \xi_2 (z, \overline z)\Big).
\ee
In this construction, the generators of $W_\infty$, $\rho [\xi ]$, are linear functionals of $\xi (z, \overline z)$ and satisfy 
at a quantum level~\cite{wcontr}:
\be 
\Big[\, \rho[\xi_1], \, \rho[\xi_2]\Big] = i \rho[\{\{ \xi_1, \, \xi_2 \}\}_{\rm Moyal}].
\ee
The classical area-preserving $w_\infty$ Lie algebra,  obtained from $W_\infty$ by an appropriate contraction discussed in \cite{wcontr},  is then
\be 
\Big[\, \rho[\xi_1], \, \rho[\xi_2]\Big] = i \rho[\{ \xi_1, \, \xi_2 \}_{\rm Poisson}],
\ee
where $\{., \, .\}_{\rm Poisson}$ denotes the (classical) Poisson bracket. 

Notice that, in this representation, the $W_\infty$ gauge fields $A_\mu(x, z, \overline z)$ are defined in a $d+2$-dimensional space time
$\{x,\, z, \,\overline z\}$ with a two-dimensional `internal' space spanned by the $\{z, \, \overline z\}$ coordinates. 
The Yang-Mills-type $\mathcal S$ action, which is invariant  under the $W_\infty$ gauge transformations, has the form
(\ref{gtr}),  
\begin{equation}\label{YMahat}
\mathcal S = -\frac{1}{4g^2} \, \int d^d x \frac{1}{4} {\rm Tr} \Big(\hat \mathcal F_{\mu\nu}\, \hat \mathcal F^{\mu\nu} \Big)~:~
\hat \mathcal F_{\mu\nu} = 
\partial_\mu \hat A_\nu (x) - \partial_\nu \hat A_\mu (x)  -i  \Big[\hat A_\mu, \, \hat A_\nu \Big]~, 
\end{equation}
where $g$ is a coupling constant, and can be rewritten  using the coherent-state representation as~\cite{wcontr}:
\bea\label{YM}
\mathcal S &=& -\frac{1}{4g^2}\,  \int d^d x d^2 z \, \sum_{n=0}^{\infty}\, \frac{(-1)^n}{n\!} \, \partial^n_z {\mathcal F}_{\mu\nu}(x, z, \overline z) \, \partial_{\overline z}^n {\mathcal F}^{\mu\nu}(x, z, \overline z)~: \nonumber \\
{\mathcal F}_{\mu\nu} &=& \partial_\mu A_\nu (x, z, \overline z) - \partial_\nu A_\mu (x, z, \overline z)  + \{\{ A_\mu, \, A_\nu \}\}_{\rm Moyal}(x, z, \overline z)~.
\eea
The reader should notice the non-local nature of the action in terms of the $z,\overline z$ variables. In fact, as stressed in \cite{wcontr},
it is this non-local nature of the action that differentiates the $W_\infty$ from $w_\infty$ as far as the association with the $SU(\infty)$ gauge theory is concerned.
It is the spectrum of the $W_\infty$ that can be viewed as the $N \to \infty$ limit of $SU(N)$, \emph{not} that of the $w_\infty$.

Thus we see that, in analogy with the $SU(N \to \infty)$ example discussed previously, the group Trace (here over the infinite-dimensional $W$-algebra) is replaced by an integral over the 
coordinates of the two-sphere $S^2$. In fact, in the action (\ref{YM}) there are no exponentially-damped $e^{-|z|^2}$ factors so,
in order to have well-defined expressions upon partial integration, one must require the fields and their derivatives to vanish at $z, \overline z \to \pm \infty $. 
For the case of the sphere, there is no boundary, and hence one is not facing this problem. One should identify the (non-compact) variables $z, \overline z$ 
with some stereographic-projection coordinates of the $S^2$. 
In the spirit of (\ref{compoisson}), then, one may expect that the infinite-dimensional $W_\infty$ gauge symmetry indeed
corresponds to an area-preserving symmetry that leaves invariant the area of $S^2$, ${\rm S\, Diff}(S^2)$. 

For the stringy black hole of interest, the above considerations apply if we
identify the gauged $W_\infty$ symmetry with the $W_{1 + \infty}$ of the string states discussed previously, 
which involves discrete delocalised states.
It has been shown in \cite{lee} that, for the asymptotically-flat two-dimensional string theory,
i.e., the $c=1$ Liouville model, for each of these states one can construct `discrete gauge states', 
at various mass string levels corresponding to tensorial gauge fields, which satisfy
the same $W_\infty$ algebra as the topological discrete states of the two-dimensional string theory~\cite{klebanov}. 
These states carry the $W_{1+\infty}$ charges and thus one can gauge the $W_\infty$ algebras by coupling them to the 
corresponding currents. In this way one may have an explicit realisation of the gauge transformations (\ref{gtr}) for the 
two-dimensional string. The generalization to the curved black-hole background is a non-trivial task, 
since one expects a mixing of various mass levels in the exactly-marginal vertex operators describing the discrete gauge states,
as with the standard topological states. 

In fact there is a much simpler formal picture that describes the situation non-perturbatively
in the string coupling~\cite{witt,wadia}. Scattering theory around the particular solution of 
string theory that describes the asymptotic state of the two-dimensional black hole, 
i.e., the $c=1$ Liouville string in a flat two-dimensional target-space background, is known to be described by a 
completely integrable (and soluble) one-dimensional (quantum-mechanical) matrix model. 
The latter is essentially a theory of free fermionic fields $\psi (t)$ interacting with an inverted harmonic oscillator potential. 
The fermions depend on the Liouville dimension, which is a `spatial' coordinate $r$
in this case. The $W_\infty$ charges in that case are described by the (infinite) set of the moments of the 
energy (Hamiltonian). In fact, if one considers a state with incoming fermions of energies $\epsilon_i$, $i=1, \dots k$,
the conserved charges are
\be\label{charges}
Q_n = \sum_{i=1}^k \epsilon_i^n  ~,
\ee
which are conserved for each $n=1, 2 \dots \infty$, where the charge $n=1$ is the Hamiltonian of the system. 

The $W_\infty$ symmetry algebra associated with these charges has a classical limit
that is associated with the canonical transformations that preserve the
free-fermion phase-space (area-preserving phase-space form), as we have discussed previously. 
From the detailed string theory considerations of the associated Liouville theory described previously, 
we know that there exist discrete gauge states that carry these charges~\cite{lee,wadia},
i.e., the $Q_n$ (\ref{charges}) can couple to gauge fields, which constitute elements of  the gauged $W_{1+\infty}$ 
Lie algebras discussed above, leading to the $w$ hair of the black hole~\cite{emn1}. 
In the context of the matrix model, the space-time dimension of the gauge field is $d=1$. 
The embedding of such matrix models in four dimensions can then be done by identifying the 
internal colour space $z, \overline z$ with the horizon surface of the four-dimensional black hole, 
whose asymptotic state is described by the c=1 Liouville string theory.

\subsection{Hawking radiation from spherically-symmetric black holes and $W_{1+\infty}$ symmetries}

We close this Section by mentioning some important results that provide 
a link between phase-space-area-preserving $W_{1+\infty}$ algebras and 
Hawking radiation in generic four-dimensional black holes with spherically symmetric horizons,
which aids understanding the connection between phase-space $W_{1+\infty}$ symmetries and
area-preserving diffeomorphisms of the horizon 
in a more generic context than the $SU(\infty)$ black hole case discussed previously. 
It was shown in~\cite{higherspin, bonora} that moments of the Hawking radiation
emitted by a generic, non-stringy spherically-symmetric black hole are
connected to a $W_{1+\infty}$ algebra carried by higher-spin states,
whose currents are sourced by background fields of higher-spin states, which
can be identified with the discrete gauge states~\cite{lee} discussed in our stringy approach~\cite{emn1} to black-hole singularities.

Crucial to the connection of Hawking radiation to $W_\infty$ algebras is the effective two-dimensional conformal field theory representation~\cite{wilczek} of the dynamics of matter fields in the near-horizon geometry of a spherically-symmetric black hole, as adopted in our approach~\cite{emn1}. Within this context, it is known that  
the quanta of Hawking radiation emitted from the horizon of a spherically-symmetric black hole break general covariance. 
As shown in \cite{wilczek}, this symmetry is restored (in the sense that the corresponding gravitational anomalies
in the quantum gravity path integral are cancelled) in the case of a (1 + 1)-dimensional black body
at the black-hole Hawking temperature~\cite{hawk}. Thus, one can represent the 
effective two-dimensional field theory of the Hawking radiation on the black-hole horizon
as a two-dimensional field theory with an infinity of two-dimensional conformal quantum fields
with a thermal spectrum, with the left movers corresponding to radially infalling matter and right movers to outgoing matter. 

Schwarzschild black holes emit Hawking radiation with a Planck distribution 
\be\label{planckdistr}
N^\pm (\omega) = \frac{1}{e^{\beta \, \omega} \pm 1}
\ee
where $\beta$ is the Hawking temperature~\cite{hawk}, $\omega$ is the frequency (energy) of the radiation quantum, and + (-) corresponds to fermions (bosons) respectively. The full spectrum of the (Fourier-analyzed) radiation is encoded in the 
higher moments or \emph{fluxes}~\cite{higherspin}. The energy flux, for instance, is given by the second moment  of $N^\pm (\omega)$, $F_2 (\omega) = \int_0^\infty \frac{d\omega}{2\pi} \omega \, N^\pm (\omega) $. The complete thermal
Hawking radiation spectrum is specified by the infinity of higher moments:
\be\label{highermomplus}
F_{2n}^+ = \int_0^\infty \frac{d\omega}{2\pi} \omega^{n-1} \, N^+ (\omega) = (1 - 2^{1-2n})\, \frac{B_{2n}}{8\pi \, n}\, \kappa^{2n}~,
\ee
or
\be\label{highermomminus}
F_{2n}^- = \int_0^\infty \frac{d\omega}{2\pi} \omega^{n-1} \, N^- (\omega) =  \frac{B_{2n}}{8\pi \, n}\, \kappa^{2n}~,
\ee
where the $B_{2n}$ are the Bernoulli numbers and $\kappa = 2\pi/\beta $ is the surface gravity of the black hole. 

An interesting proposal was made in~\cite{higherspin} that the higher fluxes $F_{2n}$, $n > 1$
could be connected to phenomenological higher-spin currents (with appropriate normal ordering), 
\emph{i.e.} higher-spin generalizations of the energy-momentum tensor, 
which is the spin-two current associated with the energy flux.  As shown in \cite{higherspin}, 
these higher currents can be expressed in terms of two dimensional boson and fermion fields, 
involving linear combinations of (two-dimensional)
space-time derivatives acting on the fields. In an  appropriate light-cone coordinate ($u, v$) description, 
the outgoing near-horizon Hawking radiation is described by holomorphic ($u$-dependent) currents of the
following form in the scalar case~\cite{higherspin}:
\be\label{holomorphic}
J^B_{uu\dots u} = {\rm linear~combinations~of} \, : (-1)^{n + m} \, \partial_{u}^m \, \phi \, \partial_u^{2n - m} \, \phi :~,
\ee
where $: \dots :$ denotes appropriate normal ordering, as defined in \cite{higherspin}~\footnote{For
the fermion case, see~\cite{higherspin} and~\cite{emn2016}.}.

It should be noted that there are ambiguities in the representation of the higher moments of the 
Hawking radiation in terms of conformal fields on the horizon,
reflected in the relative coefficients of the various terms appearing in the holomorphic currents (\ref{holomorphic}).
Moreover, the currents are not normalizable in general.
However, these issues have been resolved by the proposal of \cite{bonora} that the coefficients of these currents
be fixed by a symmetry principle, namely 
by the postulate that there is a higher symmetry in the horizon of the black hole than the Virasoro symmetry, namely a $W_\infty$ algebra. 
It was to be expected from their higher-derivative holomorphic structure, $\partial^n_z \chi $ ($\chi=\phi, \psi$)~\cite{iso}
that the currents could be cast in such a way as to form an infinite-dimensional algebra. 

In a flat two-dimensional space-time, after \emph{Euclideanisation} and replacing the light-cone coordinates $u, v$ by the 
complex variables $z$, $\overline z$ respectively, the $w_\infty$-generating bosonic currents for the conformal spin $s$ states
can be written in the form 
\be\label{wbosonic}
j_{z\dots z}^{(s) B} = q^{s-2}\frac{2^{s-3}\, s\, !}{(2s-3)!\,!}\, \sum_{k=1}^{s-1} \, (-1)^k \, \Big[\frac{1}{s-1}\, \begin{pmatrix} s-1 \\ k \end{pmatrix} \, \begin{pmatrix} s-1\\ s-k \end{pmatrix} \, \Big] \, : \, \partial_z^k \phi(z) \, \partial_z^{s-k} \overline{\phi}(z) \, ,
\ee
where $:\dots : $ denotes normal ordering, defined in \cite{bonora}, and $q$ is a complex (in general) 
deformation parameter~\cite{iso}. In this formalism the holomorphic free fields $\phi(z)$ are assumed to have
two-point functions of the form $\langle \phi(z) \, {\overline \phi}(z^\prime) \rangle = -{\rm ln} (z-z^\prime)$, 
with the others vanishing. The deformation parameter $q$ can be fixed~\cite{bonora} by demanding that the 
currents (\ref{wbosonic}), when \emph{covariantised} as appropriate for the
curved space-time of the spherical symmetric black hole (which is only conformally equivalent to a flat space),
reproduce the higher moments of the Hawking flux. 
The $s=2$ current is independent of the $q$ deformation parameter, as expected from
the fact that this current can be identified unambiguously with the holomoprphic stress tensor
$$ j_{uu}^{(2)} = -2\pi \, T^{\rm hol}_{uu} ~.$$
However, the higher-spin currents depend on $q$, and one must fix $q=-i/4$~\cite{bonora}
in order to reproduce the Hawking fluxes  for bosons (\ref{highermomminus}) when 
covariantising the expression (\ref{wbosonic}) by replacing the 
ordinary derivatives with covariant ones, as is appropriate for discussing the Hawking flux in the 
conformally-flat metric that represents the near-horizon geometry of the black hole.

As discussed in great detail in \cite{bonora}, the currents  (\ref{wbosonic}) and
their fermionic counterparts with spins higher than two are free of conformal or diffeomporphism anomalies
(or, if the latter exist, they are trivial). This is consistent with the fact that the higher 
moments of the Hawking radiation are expected to describe a theory free of gravitational anomalies,
since only the spin-two current (stress tensor) of the theory has diffeomorphism or conformal anomalies,
and their cancellation required the appearance of the Hawking radiation spectrum~\cite{wilczek}. 
Moreover, this is also consistent with the fact that, if these currents had conformal anomalies, they
would correspond to new (non-gauge) quantum numbers of the black hole, in violation of
the no-hair theorem. Hawking radiation is consistent with this theorem, which is then reflected 
in the absence of anomalies in the currents corresponding to currents with spins higher than two.  

The covariant higher-spin currents $J^{(s) B,F}_{\mu_1 \mu_2 \dots \mu_n} $ are sourced by 
appropriate background fields ${\mathcal B}^{(s)B,F}_{\mu_1\mu_2\dots \mu_n}$:
\be\label{sources}
J^{(s)B,F}_{\mu_1 \dots \mu_{n}}  = \frac{1}{\sqrt{g}}\, \frac{\delta}{\delta {\mathcal B}^{(s)B,F\, \mu_1 \dots \mu_n} }S \, ,
\ee
where $S$ is the two-dimensional effective action of the Hawking radiation in the near-horizon geometry of the 
spherically-symmetric black hole. The relevant interactions in this effective geometry are then simply given by
\be\label{e2dg}
S_{\rm int} = \int_{\rm near~horizon~2D~space-time} \, d^2 x \sqrt{g}  \, \sum_{\alpha=B,F} {\mathcal B}^{(s) \alpha\, \mu_1 \dots \mu_n} \, 
J^{(s) \alpha}_{\mu_1 \dots \mu_{n}}~,
\ee
and the background fields ${\mathcal B}^{(s) \alpha \, \mu_1 \dots \mu_n}$ may be taken taken to vanish at 
asymptotic spatial infinity. Eqn.~(\ref{e2dg}) generalises the spin-2 case,
in which the corresponding spin-2 current (the stress tensor) couples to the graviton
field, $\int d^2 x \sqrt{g} \, T^{\mu\nu} \, g_{\mu\nu}$, which is characterised by diffeomorphism invariance 
(acting as a `gauge symmetry'):  $\delta g_{\mu\nu} = \partial_{(\mu} \xi_{\nu)}$ for an infinitesimal 
diffeomorphism $\xi_\mu \to x_\mu + \xi_\mu$, provided the stress tensor is 
conserved~\footnote{In the black-hole case, as we have discussed above, 
the diffeomorphism invariance is broken by the outgoing flux, but the form of the transformation is included in (\ref{gs}).}.
As already remarked, the higher-spin currents are free from conformal and diffeomorphism anomalies~\cite{bonora}
and are conserved exactly, and their conservation is associated with 
an infinity of Abelian gauge symmetries of the form
\be\label{gs}
{\mathcal B}_{\mu_1 \dots \mu_n}^{(s)} \rightarrow {\mathcal B}_{\mu_1 \dots \mu_n}^{(s)}  + \partial_{(\mu_1 } \, \Xi_{\mu_2 \dots \mu_n)} \, ,
\ee
where the $(\dots )$ indicate the appropriate symmetrization of indices. The presence of these infinite gauge symmetries is 
consistent with the no-hair theorem, as the spatial integrals of the currents correspond to conserved charges. 
The existence of a $W_\infty$ symmetry of the matter in the near-horizon geometry, which is larger than
the Virasoro algebra, results in the complete integrability of the matter system, and is analogous to the case
of matter in the near-horizon geometries of black-hole structures in the context of string theory~\cite{emn1,zanon}, discussed above. 

This $W_\infty$ algebra is \emph{phase-space-preserving}, as are
the $W_\infty$ algebras discussed in the stringy cases above. To see this, one may rewrite the (traceless) 
energy-momentum tensor of the two-dimensional effective scalar theory using a point-splitting method~\cite{bonora}: 
\bea\label{tmn}
T_{\mu\nu} &=& {\rm lim}_{y \to 0}\, \partial_\mu \phi (x - y) \partial_\nu \phi (x + y) - g_{\mu\nu}\,  \Big({\rm stress-tensor \, trace} \Big)  \nonumber \\
&=& \sum_{i=0} \, \sum_{j=0} \frac{(-1)^i}{i!\, j!} : y^{\mu_1}\dots y^{\mu_i} \, y^{\nu_1} \dots y^{\nu_j} \, \partial_\mu\, \partial_{\mu_1} \dots \partial_{\mu_i} \phi (x) \, 
\partial_\nu\, \partial_{\nu_1} \dots \partial_{\nu_j} \phi (x) :
\eea
This expression is covariantised by replacing the partial derivatives by covariant derivatives, 
giving the right-hand-side of (\ref{tmn}) a complicated expansion in terms of products of the
higher-spin currents with $y$-dependent factors (\ref{tmn}) that correspond via complicated 
background tensors $\mathcal B^{(s)}_{\mu_1 \dots \mu_{i_n}}$ to the aforementioned background 
fields that source the higher-spin currents. For our purposes, the most important feature of (\ref{tmn}) 
is the fact that the right-hand side depends not only on the coordinate $x^\mu$ but also on the
quantity $y^\mu=dx^\mu$ in the cotangent bundle, and thus lives in a symplectic
phase-space manifold, showing that the $W_{1+\infty}$ algebra generated by the higher-spin
currents of the Hawking radiation spectrum is a phase-space algebra. 

One may associate these symmetries classically with \emph{horizon-area-preserving diffeomorphisms}, 
following the discussion of Section~\ref{sec:suinfty}  for the SU($\infty$) coloured black hole case, but
we reserve details for a future publication~\cite{emn2016}. For our purposes here we only mention that the 
above-mentioned $W_{1+\infty}$ field theories on the black-hole horizon (for fermion or boson fields) 
may be ``gauged'', in the way discussed in~\cite{wcontr} and reviewed in Section~\ref{sec:wgauge},
by considering the extension of the fields in a space augmented with two extra dimensions $\xi, \overline \xi $ 
that can be taken to be the coordinates of the spherical horizon surface. The corresponding field theories of these
$(d + 2)$-dimensional fields, where $d$ the target-space dimensionality of the field theories on the horizon,
are gauged in the way discussed in detail in \cite{wcontr}~\footnote{One has $d=1$ for the holomorphic fields that 
represent the outgoing Hawking radiation~\cite{wilczek,higherspin,bonora}, which in the case of string-theory-inspired 
black holes~\cite{emn1} can be represented non-perturbatively via the $c=1$ matrix model~\cite{wadia}.} 
and can be shown to be invariant under the phase-space-area-preserving
diffeomorphisms of the coordinates $\xi, \overline \xi $, which can be taken to represent the spherical black-hole horizon.
It is in this way that the infinite-dimensional dynamical phase-space-area-preserving $W_{1+\infty}$ symmetries 
(\ref{tmn}) of the near-horizon currents representing the Hawking radiation spectrum are related to classical horizon-preserving 
$w_\infty$ symmetry algebras. This is consistent with the identification of the classical black hole entropy
(that is proportional to the area) with a classically-conserved Noether charge~\cite{wald}. 

Before closing this discussion, we note an important difference of the two-dimensional $c=1$ string theory 
description~\cite{witt,Chlyk} of the black-hole singularity~\cite{emn1,emn2} from a generic two-dimensional 
field-theoretic representation of the outgoing Hawking radiation, as discussed above. In the latter case, 
as we have seen, higher-spin currents consist of higher derivatives of propagating local (boson or fermion) fields,
whereas in the $c=1$ string theory  representation one encounters \emph{necessarily} non-propagating discrete 
delocalised states in the excitation spectrum, which are non-thermal. These states exist over
and above the ordinary Hawking radiation fields and, as we have discussed above, their presence has highly 
non-trivial consequences. These delocalised states correspond to gauge states in the $c=1$ string theory~\cite{lee}, 
which carry the phase-space-area-conserving and coherence-preserving $W_\infty$ charges of the two-dimensional black hole singularity. 

In our treatment, the singularity of the four-dimensional spherically-symmetric stringy black hole is 
characterised by an infinite-dimensional $W_{1+\infty}$ phase-space symmetry, 
whose charges are carried by the (infinity of) delocalised discrete, non-propagating higher-spin gauge states that are
responsible for maintaining quantum coherence~\cite{emn1}. On the other hand, the black-hole horizon carries another 
set of phase-space $W_{1+\infty}$ symmetries, corresponding to higher-spin currents composed of the propagating modes of the 
two-dimensional effective field theory in the near-horizon geometry, representing the outgoing Hawking radiation flux. 
Thus, the Hawking radiation $W_\infty$ can be made to preserve the horizon area, but it is the 
discrete non-propagating string states  that preserve quantum coherence. 

\section{{\it Quo Vadis} Supertranslations?}

\subsection{D-Brane Recoil and Supertranslations\label{sec:dfoam} }

In order to understand better the microscopic mechanism for encoding information on the horizon,
we consider the back-reaction of the black-hole horizon induced by its interaction with infalling matter. 
We first concentrate on the two-dimensional stringy black hole where the horizon is a point in space. 
As discussed in the previous section, we can represent the horizon of such a black hole as a 
D-particle defect in space, whereas the horizon of a four-dimensional black hole can be represented by a 
spherical Dirichlet brane. Using D-branes enables us to consider the `momentary'
capture of string matter by the horizon, in the sense of a spontaneous change of world-sheet boundary conditions 
from Neuman to Dirichlet. Such a process will lead to splitting of a closed-string state into two open ones. 
If one considers closed-string states as corresponding to gravitons and open-string states to 
gauge particles (including photons), this process may correspond to the conversion of an infalling
graviton into a pair of photons. In general, when one represents the horizon of a black hole as a D-brane, 
the interaction with infalling string states implies a `recoil' of the D-brane.

In the pilot case of a black hole in two target-space-time dimensions, we consider a matter particle
represented by an open string falling into the horizon
the interaction of the string with the D-particle horizon implies that
at least one end of the open string attaches to the D-particle defect. 
As a result of the interaction, the D-particle undergoes a non-trivial change in velocity
\begin{equation}
u_r  \; = \; \frac{g_{s}}{M_{s}}\Delta p_{r} \; = \; \frac{g_{s}}{M_{s}}\xi_{r}\, p_{r} \, , 
\label{recoilv}
\end{equation}
where $\xi_{r}$ denotes the
fraction of the incident matter particle momentum that corresponds to the
momentum transfer $\Delta p_{r}$ during the scattering. 
As discussed in ~\cite{EllisDbrane} the non-trivial
capture and splitting of the open string during its interaction with
the D-particle, and the recoil of the latter, result in a \emph{local}
effective metric distortion of the form 
\begin{equation}
ds^{2}=g_{\mu\nu}dx^{\mu}dx^{\nu}=(\eta_{\mu\nu}+h_{\mu\nu})dx^{\mu}dx^{\nu}~: \qquad h_{0r}= u_{r}~.\label{recmetric}
\end{equation}
In the black hole case of~\cite{witt} with dilaton hair: $\Phi = -2 \, {\rm ln}{\rm cosh} \, r$, 
the string coupling $g_s = e^\Phi$ becomes weak at large distances: $g_s \to 0$ for $r \to \infty$. 
Hence $u_i \to 0$ and the space-time distortion vanishes at large distances, where the space is asymptotically flat.

The metric (\ref{recmetric}) can be generalised to higher dimensions, with
a D-brane horizon recoiling along the $i$'th spatial dimension, in which case the space-time distortion due to the recoiling 
D-brane horizon can be written as
\begin{equation}\label{waterfall}
ds^2 = dt^2 + 2 u_i dx^i dt - \delta_{ij} dx^i dx^j ~.
\end{equation}
This metric was determined from world-sheet (logarithmic) conformal field theory considerations in~\cite{EllisDbrane,kogan}:
the world-sheet deformations representing the recoil of the D-brane close a logarithmic conformal algebra
on the world-sheet of the string that represents a dragging of the 
reference frame by the D-brane horizon as it moves slowly
on the flat space-time background. On the other hand, the string excitations represent relativistic particles, 
and thus they move at the local speed of light.
One may perform a time coordinate change in the metric (\ref{waterfall}) to write it in the following form, 
which is valid up to terms $u^3$ for small recoil velocities $|\vec u | \ll 1$, 
as is appropriate for macroscopic black holes:
\begin{equation}\label{waterfall2}
ds^2 = dt_{\rm ff} ^2 + 2 u_i dx^i dt_{\rm ff} - \delta_{ij} (dx^i - u^i dt_{\rm ff} ) (dx^j - u^j dt_{\rm ff}) + \mathcal{O}(u^3) \, .
\end{equation}
The metric (\ref{waterfall2}) is the Gullstrand-Painlev\'e metric~\cite{gpm}, 
which represents the geometry around the exterior of a Schwarzschild black hole. It represents the
space falling into the black hole as a Gallilean `river' on a flat space-time in which relativistic `fish' may swim. 
The river represents the frame of the recoiling D-particle horizon, while the fish are the relativistic matter strings~\cite{ms}. 
In (\ref{waterfall2}), $t_{\rm ff}$ is the time of a free-floating observer who is at rest at infinity,
compared to the centre of the black hole. We stress that, in the case of a black hole,
the relative velocities $u^i$ are coordinate-dependent, as already mentioned, 
due to the variation in the string coupling from being strong near the black hole singularity to being weak on the horizon. 

In general, the recoil velocity has components normal and tangential to the horizon.
The former can be associated with changes in the horizon area, and hence the black-hole
entropy, whereas the latter would not change the area. 
In case of such a tangential recoil component, metrics of the form (\ref{waterfall2}), written in Bondi coordinates, 
are of the same form as metrics that have been discussed in the past in connection with 
gravitational wave radiation in asymptotically-flat regions of space time~\cite{bms,bms2},
and are known to be associated with supertranslations of the Bondi retarded time $u \equiv t_{\rm ff} + r $ 
\be\label{supertransl}
u \to u + \alpha (\theta,\phi) \, ,
\ee 
where $\alpha (\theta,\phi)$ is a function of the angular coordinates $\theta, \phi$.
Such a retarded time was used in \cite{bms,bms2} to discuss outgoing gravitational wave signals 
arriving at a distant observation point. 
Such BBMS$^+$ transformations form an infinite-dimensional set of diffeomorphisms that include 
as a subgroup the four-parameter group of ordinary translations. 
In the case of matter falling into the black-hole horizon one may use instead the BBMS$^-$ 
transformations pertaining to the advanced time $v = t_{\rm ff} - r $, which amount to the supertranslations 
\be\label{supertransl2}
v \to v + \xi (\theta,\phi) \, ,
\ee 
where $\xi (\theta,\phi)$ is a function of the angular coordinates $\theta, \phi$ on the black-hole horizon brane.
The retarded (or advanced) time is viewed in general as a scalar function of the coordinates $u(x^\mu)$ that obeys 
$u _{,\, \mu} \, u_{,\, \nu} \, g^{\mu\nu} = 0$,  implying that the hypersurfaces $u={\rm constant}$ are light-like. 

The generic space-times on which there are supertranslations that leave invariant the boundary conditions
have the form~\cite{bms2}
\bea\label{bmsmetric}
ds^2 &=& (\frac{V}{r} \, e^{2\beta} )\, du^2 - 2\, e^{2\beta} du \, dr + r^2 h_{AB} \, (dx^A - U^A \, du)\, (dx^B -U^B du)~, \nonumber \\
2h_{AB}  dx^A dx^B &=& (e^{2\beta} + e^{2\gamma}) \, d\theta^2 + 4 {\rm sin}\theta \, {\rm sinh}(\gamma - \delta) \, d\theta\, d\phi + {\rm sin}^2\theta \, (e^{-2\beta} + e^{-2\gamma}) d\phi^2 \, ,
\eea
where $u$ is a retarded time and $(r, x^A)$ are the three spatial coordinates, 
with $x^A = (\theta, \phi)$ the angular variables of the four-dimensional space-time
appropriate for spherically-symmetric solutions of the gravitational equations, with
determinant ${\rm det}\, h_{AB} = {\rm sin}^2\theta$. The functions $V, U^A, \beta, \gamma, \delta$
are arbitrary functions of the coordinates and, to match them with (\ref{waterfall2}),
one performs the advanced (retarded) time transformation $u$ ($v$) 
from $t_{\rm ff}$ as mentioned previously. The asymptotic flatness of (\ref{waterfall2}), 
due to the asymptotic vanishing of the recoil-induced  distortion of the space-time surrounding the black hole horizon, 
implies the boundary conditions $ {\rm Lim}\, \Big( V/r = 1, \,  {\rm Lim }\, (r\, U^A) =  \beta=\gamma=\delta \Big) =0$
for fixed $u$ or $v$ and $r \to \infty$.

\subsection{Supertranslations are not Enough}

Hawking~\cite{hawking} has suggested that such supertranslations of an \emph{advanced} 
Bondi  time on the black-hole horizon may solve the issue of quantum coherence.
The proposal builds upon analyses in the asymptotically-flat regime of a generic Schwarzschild black hole space-time
by Strominger and collaborators~\cite{strominger,strominger2},
who have provided arguments that the horizon supertranslation transformations can be 
viewed as a conformal Kac-Moody symmetry group, entailing an infinity of conservation laws. 
This symmetry group is an infinite-dimensional diffeomorphism group that leaves invariant
the asymptotic BMS states as well as the generic quantum gravity scattering matrix
defined by means of asymptotic \emph{in} and \emph{out} quantum states. 
Section~6 of the first reference in \cite{strominger2} made an association of these infinite supertranslation
charges with hair for black holes, and suggested that such charges `may bear on the information 
puzzle'~\footnote{We know no obvious association of Kac-Moody algebras to the 
$W_\infty$ area-preserving algebras discussed here, except in the supersymmetric case considered in \cite{sorba},
where an area-preserving diffeomorphism algebra ${\rm SDiff}(\mathcal M)$ of a two-dimensional surface $\mathcal M$ acts as a derivation algebra on a
super-Kac-Moody algebra, much as Virasoro algebras act as derivation algebras on Kac-Moody algebras on a one-dimensional circle $S^1$. 
The implications of this result in our case are not obvious. However, we recall that when one supersymmetrises the two-dimensional black hole 
to a (twisted topological) $N=2$ theory, a double $W_{1+\infty} \otimes W_{1+\infty}$ describes the singularity structure~\cite{wbreak}, which is broken
$\to W_{1+\infty}$ away from it. The findings of~\cite{sorba} may be relevant in the case when this $N=2$ model is embedded in four dimensions.
However, any connection between the supertranslation U(1) Kac-Moody algebra on the black-hole horizon~\cite{strominger} with
(super-)$W_\infty$ horizon-area-preserving diffeomorphisms is currently unclear.}. 

This suggestion looks similar in nature to the one we described above and in our previous works~\cite{emn1,emn2}, 
but there are important differences, and we do not think that supertranslations are sufficient to retain quantum coherence.

The supertranslations (\ref{supertransl}), (\ref{supertransl2}) are symmetries of the asymptotic (large $r$) 
metrics and not of the full black hole background space-time, 
since they correspond to physically-inequivalent metrics, as discussed in~\cite{bms,bms2}. 
This may be viewed as a sort of `spontaneous breaking' of the supertranslation symmetries of the 
asymptotic Minkowski space-time by the black hole background. The associated Goldstone bosons 
have been identified tentatively (in the semiclassical black-hole limit of infinite entropy)~\cite{dvali}
with delocalised graviton states of infinite wavelength. This proposal may sound similar to our topological stringy states,
though the situation for a finite-area (finite-entropy) black hole is far from clear.

However, in our picture the discrete topological string states include such soft graviton
states as a subset, along with an infinity of other higher-spin discrete (topological) states, 
all corresponding to conserved charges.  These higher-spin states 
are separated by mass gaps $\propto n M_s/g_s$ , $n \in Z^+$.
Hence, given that on the horizon the string coupling is weak, such states may \emph{seem} to
be decoupled at the level of the local effective field theory level (LEFT), leaving
only the massless graviton states as relevant. However, we conjecture that, for
information retention at a \emph{finite-area} black-hole horizon, one must consider the entirety of the delocalised (topological) 
higher-spin states, whose treatment goes beyond LEFT.
These realize an infinite-dimensional \emph{area-preserving} quantum $W_\infty$
symmetry algebra (with its infinity of conserved charges) in the coset black-hole model. 
This picture is in agreement with the representation of the two-dimensional black hole horizon
in four space-time dimensions as the world-sheet of a string, where a classical $w_\infty$ world-sheet symmetry 
preserves the area, and thus ensures information retention. Such world-sheet $w_\infty$ symmetries 
can be elevated to target space by means of appropriate world-sheet deformations, 
corresponding to the various excitations of the higher-spin target-space states~\cite{wmeasure}.
Moreover, we have argued that a recoil displacement of a D-brane due to a `sudden' impulse,
as is the case of a string splitting on the horizon brane, merely mislays information
that is stored on the horizon~\cite{mislays}, in the sense that entanglement is
induced between the recoiling quantum D-brane horizon and the quantum matter subsystem. 
The precise reproduction of the finite-area black-hole entropy for the generic four-dimensional black-hole case 
remains to be worked out (although arguments have been given in this direction for the two-dimensional stringy 
black hole in \cite{emn2}), and we plan to return to this issue in the near future. 

We consider that our view of discrete W-hair and the representation of the black-hole horizon
as a D(irichlet) brane recoiling under the interaction with infalling matter
is more appropriate than supertranslation invariance for maintaining quantum coherence. 
We would argue that any approach based on general relativity alone is limited in scope,
and that string-theoretical considerations based on the rigorous counting of black-hole
microstates, such as those outlined above, constitute a much more promising approach
to the black-hole information-loss problem. In particular, we recall that the $W_\infty$
algebra corresponds to an SU($N \to \infty$) algebra, as discussed in Section~4.4, and that the superstranslations 
(and the corresponding U(1) gauge states) do not include candidates for all the
infinitely-coloured hair discussed there.

\section{Conclusions and Outlook \label{sec:concl} }

We have reviewed and extended in this paper our previous arguments for the importance 
of infinite-dimensional $W_\infty$ symmetry in information retention by stringy black
holes. This symmetry preserves two-dimensional area. As such, it plays a key r\^ole
at the world-sheet level, `balancing the information books' ~\cite{emn1} in the two-dimensional
black-hole model of~\cite{witt}, which may be elevated to four dimensions to describe a
spherically-symmetric black hole~\cite{emn2}. In this case, the entropy is proportional to the area of
the horizon, so entropy is therefore clearly conserved by a $W_\infty$ symmetry. The same
symmetry preserves the area of the two-dimensional phase space describing a fermion
interacting with a four-dimensional extremal black hole in the context of 
$N=2$, $D=4$ supergravity~\cite{zanon}.

As we have stressed in this paper, there are still many open issues in our approach to
the black-hole information problem based on $W_\infty$ symmetry, but we consider it to
be much more complete and promising than recent suggestions~\cite{strominger,strominger2} based on supertranslations~\cite{bms}.
The $W_\infty$ symmetry is firmly embedded within string theory, which must surely be
taken into account in any resolution of the black-hole information problem. Also, as can be
seen from its relation to SU($\infty$), $W_\infty$ symmetry certainly offers conserved
charges ($W$-hair) that are not provided by supertranslations.

We have argued elsewhere that the $W$-hair of a four-dimensional
black hole is in principle measurable~\cite{wmeasure}, though this may not be feasible in practice~\cite{emnQM}.
For this reason, we consider that black holes `mislay' information rather than `lose' it.
However, we still lack a specific `Ariadne's thread' of external measurements that can enable us
to reconstruct the information `mislaid' within the black-hole labyrinth.

\section*{Acknowledgement}

The authors thank George Leontaris for giving us the opportunity to meet and discuss these issues at Planck~2015, 
and to present them in these Proceedings.

\end{document}